\documentclass[a4paper,12pt,twoside]{article}
\usepackage[utf8]{inputenc}
\usepackage[paper=a4paper,left=25mm,right=25mm,top=25mm,bottom=25mm]{geometry} 
\usepackage{graphicx} % Required for inserting images
\usepackage{setspace}
\usepackage{amsfonts}

\usepackage{amsmath,amssymb,amsthm}
\usepackage{bm}
\usepackage{mathtools}

\usepackage{caption}

\newcommand*{\boldone}{\text{\usefont{U}{bbold}{m}{n}1}}

\DeclareMathOperator*{\argmin}{arg\,min}
\DeclareMathOperator*{\argmax}{arg\,max}

\def\D{\mathrm{d}}

\usepackage{algorithm2e}
\usepackage{xcolor}
\usepackage{hyperref}
\usepackage{comment}

\usepackage{multirow}

\usepackage{array}

\usepackage{orcidlink}

\usepackage{authblk}

\usepackage[
backend=biber,
style=apa,autocite=inline,
sorting=nyt
]{biblatex}

\title{A categorization of performance measures for estimated non-linear associations between an outcome and continuous predictors}

\def\correspondingauthor{\footnote{Corresponding author, e-mail: \href{mailto:daniela.dunkler@meduniwien.ac.at}{daniela.dunkler@meduniwien.ac.at}}}

\author[1]{Theresa Ullmann
\orcidlink{0000-0003-1215-8561}}
\author[1]{Georg Heinze
\orcidlink{0000-0003-1147-8491}}
\author[2]{Michal Abrahamowicz
\orcidlink{0000-0002-3172-3952}}
\author[3]{Aris Perperoglou
\orcidlink{0000-0003-1355-2800}}
\author[4]{Willi Sauerbrei
\orcidlink{0000-0002-6792-4123}}
\author[5]{Matthias Schmid
\orcidlink{0000-0002-0788-0317}}
\author[1]{Daniela Dunkler
\orcidlink{0000-0003-1339-0311}
\correspondingauthor{}}
\author[ ]{for TG2 of the STRATOS initiative}

\affil[1]{\small Institute of Clinical Biometrics, Center for Medical Data Science, Medical University of Vienna, Vienna, Austria}
\affil[2]{Department of Epidemiology \& Biostatistics, McGill University, Montreal, Canada}
\affil[3]{Statistics and Data Science Innovation Hub, GSK, Stevenage, UK}
\affil[4]{Institute of Medical Biometry and Statistics, Faculty of Medicine and Medical Center, University of Freiburg, Freiburg, Germany}
\affil[5]{Department of Medical Biometry, Informatics, and Epidemiology, University Hospital Bonn, Bonn, Germany}

\date{}

\begin{document}

\maketitle

%\section*{Article Category}
%Advanced Review

%\tableofcontents

%\section*{Conflict of interest} 
%All authors declare that they have no conflicts of interest.

\section*{Abstract}

In regression analysis, associations between continuous predictors and the outcome are often assumed to be linear. However, modeling the associations as non-linear can improve model fit. Many flexible modeling techniques, like (fractional) polynomials and spline-based approaches, are available. Such methods can be systematically compared in simulation studies, which require suitable performance measures to evaluate the accuracy of the estimated curves against the true data-generating functions. Although various measures have been proposed in the literature, no systematic overview exists so far. To fill this gap, we introduce a categorization of performance measures for evaluating estimated non-linear associations between an outcome and continuous predictors. This categorization includes many commonly used measures. The measures can not only be used in simulation studies, but also in application studies to compare different estimates to each other. We further illustrate and compare the behavior of different performance measures through some examples and a Shiny app.

\section*{Graphical Abstract}
\begin{figure}[h]
    \centering
    \caption*{Suppose there is a known true nonlinear association between X and Y (black curve), e.g., in a simulation study. We want to find the best possible approximation (colored curves) to that truth.  However, there are many measures quantifying the similarity of each estimated curve to the ground truth.}
    \includegraphics[width=\textwidth]{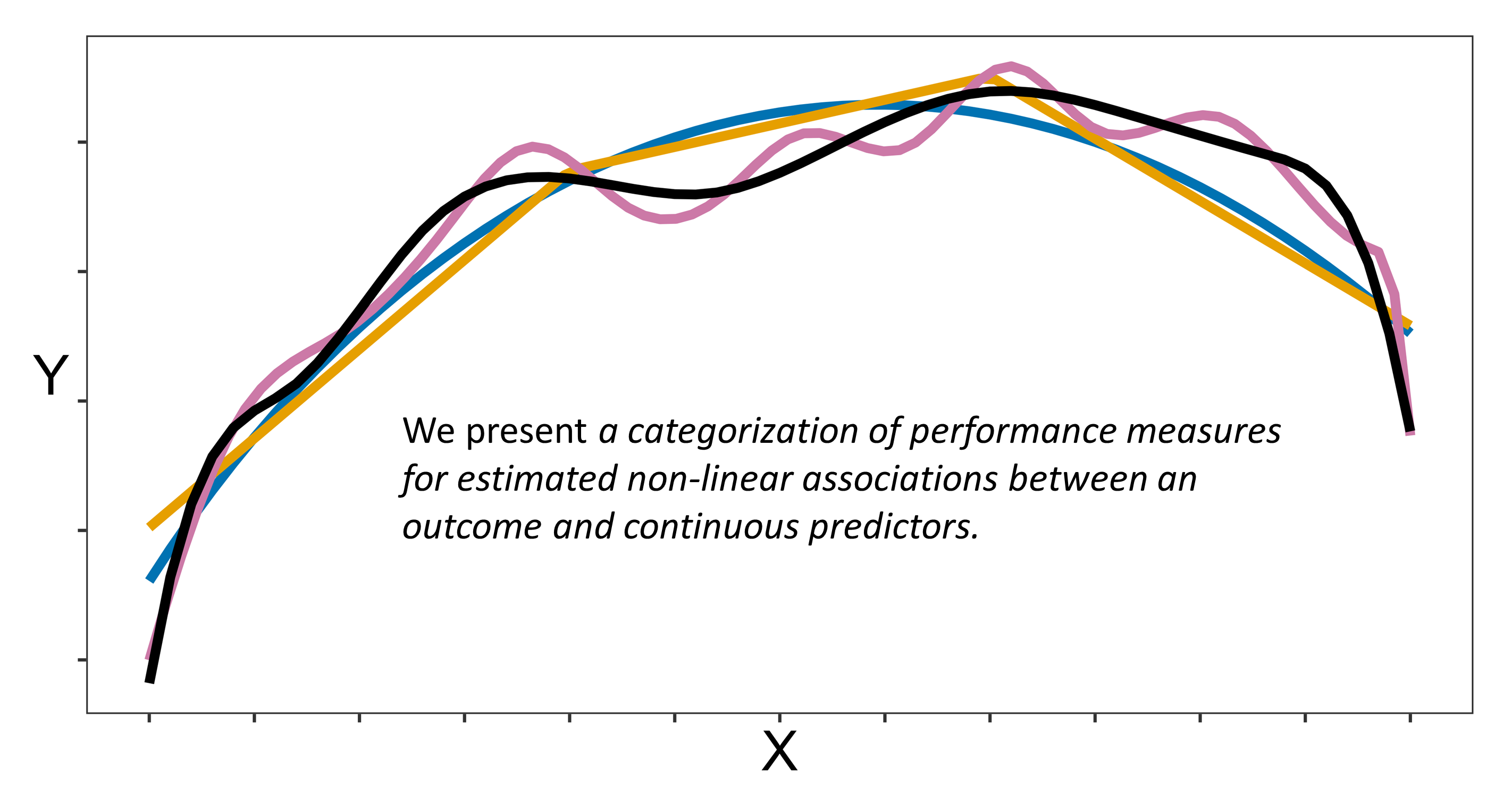}
\end{figure}

\section{Introduction}
\label{sec:intro}

\begin{figure}
    \centering
    \includegraphics[width=\textwidth]{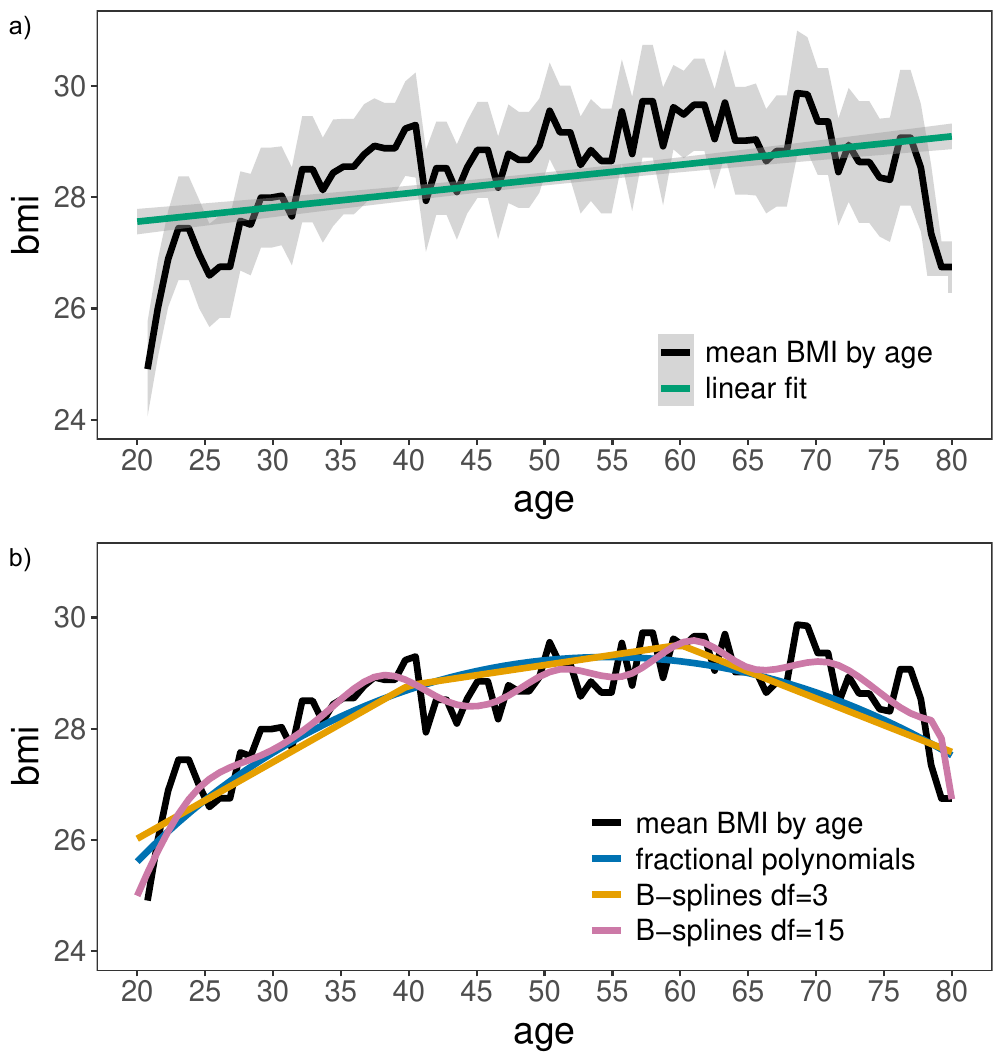}
    \caption{Association of age in years with BMI, estimated from 9377 participants of NHANES. The black line connects the mean BMI values at each age. a) The green line shows the linear fit. The shaded areas show the 95\% confidence intervals. b) Blue line: fit with fractional polynomials, orange line: fit with linear B-splines with three basis functions, pink line: fit with cubic B-splines with fifteen basis functions. For the B-spline fits, the inner knots are positioned at suitable quantiles of age.}
    \label{fig:fig1}
\end{figure}

Regression models often describe associations between an outcome and one or multiple predictors (also referred to as explanatory variables, covariates, or independent variables). Determining the functional forms of the continuous predictors in such models is an important aspect of model-building and may substantially affect the validity and interpretation of the model. Often, the associations of the continuous predictors with the outcome are simply assumed to be linear (potentially after transformation of the expected outcome via a link function). However, modeling the effect of one or multiple predictors as \textit{non-linear} can be more appropriate in many applications. Consider the example in Fig~\ref{fig:fig1}a) which shows the association of age with body mass index (BMI) based on data from 9377 participants of the National Health and Nutrition Examination Survey (NHANES) (\cite{nhanes2024}). The black curve connects the mean BMI values at each age. To account for the sampling variation reflected in the excessive wiggliness of the black curve, we can try to find a smooth model fit to better describe the systematic changes in the mean BMI values with increasing age. 
The linear fit, shown in green color, does not describe the mean BMI values well (by definition, the linear fit is unable to account for the clear non-monotonicity of the relationship), such that options for a non-linear functional form should be explored. 

There are many different methods to extend the framework of linear predictor models in order to model non-linear associations. For example, the effect of a predictor can be modeled with a polynomial function. The \textit{fractional polynomial approach} (\cite{royston2008multivariable}) extends the simple polynomial approach; here one or several power transformations are applied to a predictor. Such transformations are global, i.e., a (fractional) polynomial function is defined over the whole range of the predictor. In contrast, \textit{spline-based approaches} are local smoothers: the range of the predictor is divided into segments, a different polynomial is defined on each segment, and the polynomials (all having the same degree) are smoothly joined at the knots, i.e., the endpoints of the segments, to ensure continuity of the fitted function. Spline-based approaches include, e.g.,\ B-splines (\cite{boor1978practical}),  restricted cubic splines (\cite{stone1985additive}), P-splines (\cite{eilers1996flexible}), and thin-plate regression splines (including smoothing splines) (\cite{wood2003thin}); see \textcite{perperoglou2019review} for an overview. Fig~\ref{fig:fig1}b) again shows the NHANES example, now with a fractional polynomial fit and two different B-spline fits (linear B-splines with three basis functions and cubic B-splines with fifteen basis functions). The fitted curves have different properties: all curves reflect the aforementioned non-monotonicity, but they differ with respect to (local) extrema and the degree of wiggliness. It is not immediately clear which curve is the ``best one'', nor how one should define a criterion for determining such a ``best'' curve. 

\begin{figure}
    \centering
    \includegraphics[width=\textwidth]{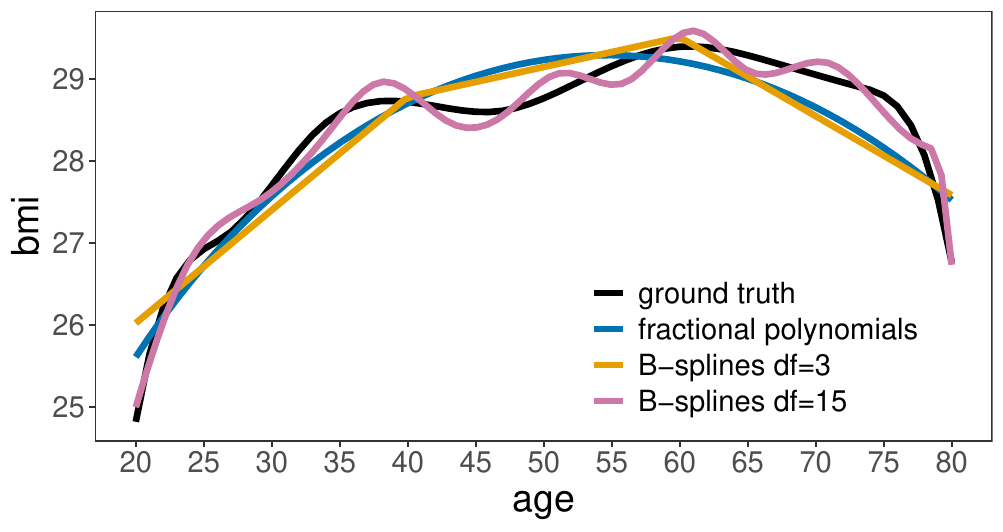}
    \caption{Illustration of a hypothetical simulation study. The black line depicts the ground truth curve. %; this ground truth is the restricted cubic splines fit. 
    The blue line shows a fit with fractional polynomials, the orange line a fit with linear B-splines with three basis functions, and the pink line a fit with cubic B-splines with fifteen basis functions. The three fitted lines can be interpreted as the curves obtained in one simulation repetition from three different methods that are compared in the simulation study. The similarity of the estimated curves with the ground truth curve can be quantified by some suitable measure, but it is not immediately clear which measure(s) should be used. Different measures are able to capture different aspects of the curves, e.g., the bias of the curves estimated with fractional polynomials or cubic B-splines with three basis functions, or the wiggliness of the curve estimated with cubic B-splines with fifteen basis functions.}
    \label{fig:fig2}
\end{figure}

Currently, there is a lack of knowledge regarding the performance of different methods for modeling functional associations. Studies comparing different methods to each other are relatively rare (\cite{strasak2011comparing, binder2013comparison}). More systematic comparison studies are needed to better understand the properties of different methods and to derive recommendations about which methods might be preferable in which application contexts, depending on the relevant characteristics of the data being analyzed and/or the complexity of the underlying true functional forms. Typically, such studies include simulated data with known true data-generating functions (``ground truths''). 
As an illustrative example, suppose that we perform a simulation study which systematically compares different flexible estimation methods. We choose a true data-generating mechanism inspired by the observed association of age on BMI in the NHANES data. This scenario is depicted in Fig~\ref{fig:fig2}. %, and that a restricted cubic splines fit (the black line) is taken as the ground truth curve. 
The black line represents the assumed ground truth curve. The fractional polynomial fit and the two B-spline fits represent the curves obtained in one simulation repetition from three methods to be compared. However, it is unclear which criteria should be used to assess their success in describing the true association of the predictor and the outcome. This illustrates a general issue: it is not clear which \textit{performance measures} should be used in systematic comparison studies, i.e., which measures should be used to compare the functional forms (linear or non-linear) estimated by the different methods to the ground truth. Several performance measures have been proposed in the literature (\cite{govindarajulu2007comparing,binder2011multivariable,buchholz2014measure}), but so far we are not aware of a systematic overview and a classification of the measures. %While \cite{govindarajulu2007comparing} focuses on non-linear exposure-response relationships using Cox models, the technical report of \cite{binder2011multivariable} is concerned with multivariable regression models with potentially non-linear effects of continuous covariates. The article by \cite{buchholz2014measure} deals with a special case, a performance measure for assessing functions of time-varying effects in survival analysis. 
In particular, it is not clear which characteristics of the estimated functional forms are addressed by different measures and how similar certain measures might be to each other.

The perspective that better comparison studies should be conducted and that more discussion is needed about suitable performance measures is shared by the STRATOS initiative (STRengthening Analytical Thinking for Observational Studies, \cite{sauerbrei2014strengthening}). The STRATOS initiative is an international consortium of biostatistics experts dedicated to offering guidance on designing and analyzing observational studies for both specialist and non-specialist audiences. Ideally, comparison studies should be \textit{neutral}, i.e., the authors should be (as a group) approximately equally familiar with all compared methods and they should not have a vested interest in one of the methods (\cite{boulesteix2013plea,boulesteix2017towards,boulesteix2024editorial}).

Considering the importance of using appropriate performance measures in (neutral) comparison studies, the present article aims to characterize and compare performance measures for estimated non-linear associations with an outcome. For this categorization, we will focus on appropriate measures in the context of \textit{descriptive modeling} (\cite{carlin2023uses}). That is, we focus on scenarios where the goal of the statistical modeling is to
%fit the curves that best separate the variation in the outcome explained by the predictors from the sampling variation,
%because one is interested in the precise forms of the relationships between the predictors and the outcome. 
identify variables with an influence on the outcome and--for continuous variables--derive functional forms which fit the data best. In addition to descriptive modeling, there are two other main objectives of statistical modeling (\cite{shmueli_explain_2010}): \textit{predictive modeling}, where the objective is to predict an outcome as accurately as possible, and \textit{explanatory modeling}, where the objective is to estimate the causal effect of a specific explanatory variable on the outcome adjusted for confounders. We refer to Section \ref{sec:discuss} for a brief discussion of performance measures in the context of predictive and explanatory modeling. 

Although our main motivation in the present article is the use of performance measures in the context of comparison studies/simulation studies in methodological research, some measures from our categorization can also be helpful in applied studies. For example, when analyzing a specific real-world dataset, researchers might apply two or more different methods to estimate non-linear associations (see, e.g.,\ \textcite{govindarajulu2007comparing}). They can then compare the results using suitable similarity measures to assess how similar they are to each other, rather than determining which result is closer to the ground truth (which is not known in applied studies). 
In the example in Fig~\ref{fig:fig1}b), suitable measures could be used to compare the different fits to each other.

The remainder of this article is structured as follows: in Section \ref{sec:categ}, we introduce our categorization of performance measures. In Section \ref{sec:shinyapp}, we illustrate the behavior of different measures based on examples which are available in a Shiny app. We conclude the article with a discussion in Section \ref{sec:discuss}.

\section{Categorization of performance measures for estimated non-linear associations}

\label{sec:categ}

\subsection{General notation} First, we introduce some notation. For observations $i = 1,\ldots,n$, we assume a data-generating mechanism of the form

\begin{equation*}
    y_i = \beta_0 + \sum_{j=1}^p f_j(x_{ij}) + \varepsilon_i
\end{equation*}
    
with continuous outcome $y_i$, predictors $x_{ij}$, $j = 1,\ldots,p$ (realizations of random variables $X_1,\ldots,X_p$), and error term $\varepsilon_i$. The notation easily generalizes to other outcome types by using a suitable link function and modifying the error term. While the model may contain categorical predictors in the form of dummy variables, from now on we only consider continuous predictors  without loss of generality. Some of the functions $f_j$ might be linear, i.e., $f_j(x) = \beta_j x$, while others could be non-linear.

The fitted model is denoted as

\begin{equation*}
    \hat y_i = \hat\beta_0 + \sum_{j=1}^p \hat f_j(x_{ij}).
\end{equation*}

Again, some of the estimated functions $\hat f_j$ might be linear, i.e., $\hat f_j(x) = \hat \beta_j x$, while others might be non-linear. 
Note that for some covariates $j$, the fitted function $\hat f_j$ may be linear while the true function $f_j$ is non-linear, and vice versa. %The sets of indices $\{j: f_j~\textrm{is linear} \}$ and $\{j: \hat f_j ~\textrm{is linear}\}$ are not necessarily equal. 

In the following, we assume that any type of model selection (if performed) is done, such that the fitted model(s) under consideration are fixed. Moreover, we only consider ``univariable'' performance measures in the sense that the estimated curve of each predictor of interest is evaluated individually. More precisely, the estimated curve $\hat f_j$ for each predictor $X_j$ of interest is compared to the ``true'' curve $f_j$. For notational simplicity, we will assume that the model is univariable with only one continuous predictor ($p = 1$), which allows us to write $f = f_1$, $\hat f = \hat f_1$ and $X = X_{1}$. The predictor $X$ takes values in a space $\mathcal{X} \subseteq \mathbb{R}$. $P_X$ is the probability distribution of $X$, $F_X$ is the cumulative distribution function, and $F_X^{-1}$ the corresponding quantile function. The formulas for the performance measures given further below also apply to multivariable models, by considering each predictor of interest separately and setting $f = f_j$, $\hat f = \hat f_j$, $X = X_j$ and so on for each index $j \in \{ 1,\ldots,p\}$ of interest in turn. In that case, each estimate $\hat f = \hat f_j$ is adjusted for one or more other variables, i.e., the estimated curves of the other predictors in the model.

\subsection{Description of the categorization} The general idea of our categorization of performance measures is that the measures can be described by different aspects. Consider, for example, the following measure:

\begin{equation}\label{eq:measexample}
    \int_{\mathcal{X}} |\hat f(x) - f(x)| \D x.
\end{equation}

This measure integrates the absolute difference between the estimate $\hat f$ and the true $f$ over $\mathcal{X}$. There are several characteristics of this measure. For example, the integral is used for summarizing the absolute difference between $\hat f$ and $f$, but different options for aggregation would be possible (e.g., calculating the integral with respect to $\D F_X$ instead of $\D x$, or taking the maximum absolute difference over $\mathcal{X}$). The measure aggregates the information across the entire range $\mathcal{X}$. Alternatives are possible, e.g., restricting the range to a subset of $\mathcal{X}$ (such as the range between the 5\% and 95\% quantile of $F_X$), with the extreme case that the difference between $\hat f$ and $f$ is only evaluated at a specific point $x^*\in \mathcal{X}$. Moreover, instead of the absolute loss, other types of losses might be considered. Finally, instead of comparing $\hat f$ with  $f$, one might also wish to compare the derivatives $\hat f'$ with $f'$ and/or $\hat f''$ with $ f''$. 

\begin{figure}[h]
    \centering
    \includegraphics[width=1\textwidth]{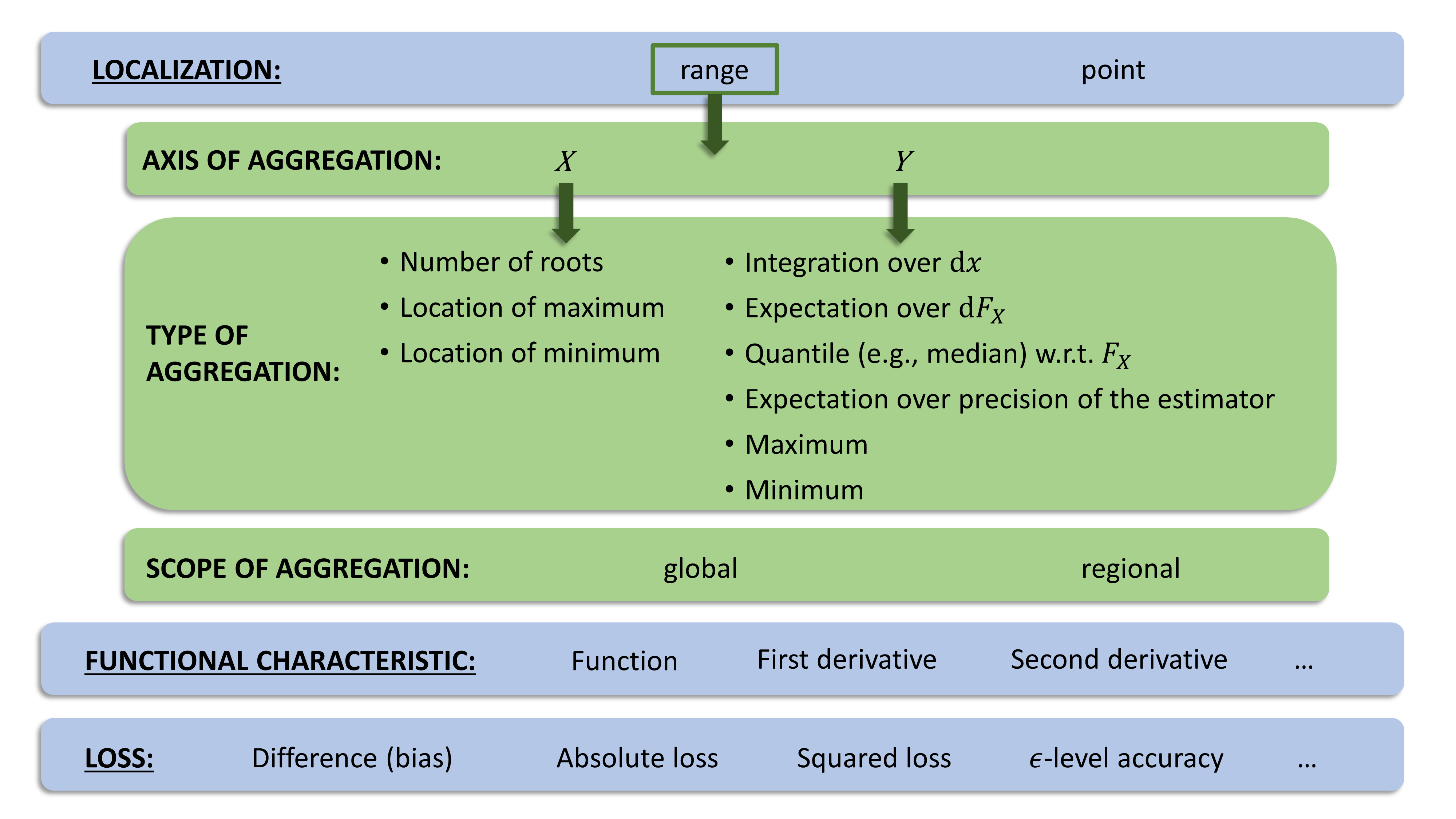}
    \caption{Categorization of performance measures. Each performance measures can be obtained by choosing specific options for each of the aspects (localization, functional characteristic, etc.).}
    \label{fig:fig3}
\end{figure}

This example demonstrates that there are several aspects of performance measures and that different performance measures can be defined by specific combinations of these aspects. This leads to our proposed categorization of performance measures as depicted in a schematic overview in Fig~\ref{fig:fig3}. First, we describe the three main aspects shown in the blue panels. 
\begin{itemize}
    \item The aspect \textit{localization} distinguishes between measures that aggregate information over a \textit{range} (there are different options for the scope of aggregation, as described below) and measures that only take performance at a specific \textit{point} $x^* \in \mathcal{X}$ into account. 
    \item We can also categorize performance measures based on \textit{the functional characteristic}, i.e., whether they evaluate the \textit{function}, the \textit{first derivative of the function}, or the \textit{second derivative of the function}, provided that these derivatives exist. When the functional characteristic is the first derivative, the slopes of the curves are compared; for the second derivative, the ``wiggliness'' of the curves is compared (\cite{green1993nonparametric}). While the third, fourth etc.\ derivatives could also be considered, these are generally more difficult to interpret and are thus not commonly used in practice. 
    \item  Finally, different types of \textit{losses} can be chosen to measure the deviation of the estimated curve from the true curve: \textit{difference loss (bias), absolute loss, squared loss,} or \textit{$\epsilon$-level accuracy}. The $\epsilon$-level accuracy (\cite{hirji1989median}) evaluates in a binary way whether two values are ``close'' to each other, and is defined as $\boldone_{\{|z_1 - z_2| \leq \epsilon\}}$ for two real numbers $z_1, z_2$, with $\epsilon > 0$ a suitable (small) value, 
    %For example, $\boldone_{\{|0.8 - 0.4| \leq 0.05\}} = 1$ (the values are close w.r.t.\ $\epsilon = 0.05$), and $\boldone_{\{|0.9 - 0.3| \leq 0.05\}} = 0$ (the values are not close). 
    see below for further details.
    We consider these four types of losses due to their popularity; other loss functions could, of course, also be used.
\end{itemize}

For measures that aggregate the information over a \textit{range}, there are some further aspects to consider (panels in green color in Fig~\ref{fig:fig3}): 
\begin{itemize}
    \item Most measures quantify the deviation from the truth on the \textit{Y axis}. For this purpose, there are different \textit{types of aggregation}: integration over $\D x$, expectation over $\D F_X$, a quantile (e.g., the median) with respect to $F_X$, expectation over the precision of the estimator, maximum, or minimum. 
    \item Other measures compare the curves on the \textit{X axis}, i.e., how the estimated curve and the true curve differ from each other with respect to the number of roots (i.e., the points where the functions or their derivatives take the value zero), the location of the global maximum in the range, or the location of the global minimum in the range. 
    \item For both types of performance measures (axis of aggregation = ``Y'' and axis of aggregation =``X''), the scope of aggregation has to be specified. \textit{Global} measures consider the information on the whole range $\mathcal{X}$ of $X$, while \textit{regional} measures consider a subrange of $X$. A typical subrange is the range between a lower quantile $l$ and an upper quantile $u$ of the distribution $F_X$ where the variable often takes values (e.g., the 5\% quantile $F_X^{-1}(0.05)$ and the 95\% quantile $F_X^{-1}(0.95)$).  
Here, the motivation is that the behavior of the curves might be less relevant in the boundaries of the range, i.e., beyond $l$ and $u$. Restricting the range might also be necessary to obtain well-defined and finite values for the measure, see below for more details. While the range between a lower and an upper quantile appears to be the most common type of subset considered in the literature (\cite{binder2011multivariable,buchholz2014measure}), the scope of aggregation could in principle also be another kind of subset of $\mathcal{X}$. For example, in medical applications, the region above a critical threshold of a clinical predictor may be of particular interest. 
%In any case, for the purposes of our categorization, we call measures ``global'' if they compare the curves at more than a single point, which means that the loss must be aggregated in some way. This distinguishes global from local measures; for the latter, aggregation is not necessary.  
\end{itemize} 

\subsection{Some examples of measures} Each performance measure arises from choosing specific options for each of the aspects stated above. For example, the measure given in (\ref{eq:measexample}) is the measure with localization = ``range'', axis of aggregation = ``Y'', type of aggregation = ``integration over $\D x$'', scope of aggregation = $\mathcal{X}$, functional characteristic = ``function'', and loss = ``absolute''. If the type of aggregation is the expectation over $\D F_X$ instead, we obtain the following measure:

\begin{equation}\label{eq:measexample2}
    \int_{\mathcal{X}} |\hat f(x) - f(x)| \D F_X(x)
\end{equation}

Compared to measure (\ref{eq:measexample}), measure (\ref{eq:measexample2}) gives more weight to the areas with more data, and less weight to the areas where the data is sparse. 

As another example, the following measure results from setting localization = ``range'', axis of aggregation = ``X'', type of aggregation = ``location of maximum'', scope of aggregation = ``global'', functional characteristic = ``first derivative'', and loss = ``absolute'':

\begin{equation}\label{eq:measexample3}
    |\argmax_{x \in \mathcal{X}} \hat f'(x) - \argmax_{x \in \mathcal{X}} f'(x)|
\end{equation}

This measure compares the locations of the maxima of the first derivatives, i.e., the positions of the steepest ascents of $\hat f$ versus $f$. Suppose we are less interested in the absolute difference of the locations of the maxima per se, but only in whether the maxima are located within some distance $\epsilon$ of each other. If so, then switching from the absolute loss to the $\epsilon$-level accuracy results in the measure

\begin{equation}\label{eq:measexample4}
    \boldone_{\{ |\argmax_{x \in \mathcal{X}} \hat f'(x) - \argmax_{x \in \mathcal{X}} f'(x)| \leq \epsilon\}}
\end{equation}

which evaluates whether the positions of the steepest ascents of $\hat f$ and $f$ are sufficiently close. For all measures involving the $\epsilon$-level accuracy, suitable values for $\epsilon$ have to be chosen. This depends on how ``strict'' a researcher wishes the measure to be, as well as on the domain of $f$ (if axis of aggregation = ``X'') or the range of $f$ (if axis of aggregation = ``Y''). For example, suppose that $f$ is defined on the domain $[0,1]$. Then for a measure as in (\ref{eq:measexample4}), where axis of aggregation = ``X'', e.g.\ $\epsilon = 0.05$ could be a reasonable choice (evaluating whether the positions of steepest ascents differ by at most 0.05). Generally, one could choose $\epsilon$, e.g., as 0.05 times the length of the domain/range of $f$, if the domain/range is a bounded interval, or 0.05 times the length of the interval between suitable quantiles of $F_X$ (as described above) if the domain of $f$ is unbounded. 

The examples of measures given so far all have localization = ``range''. In contrast, point-specific measures compare the estimated curve and the true curve at a single point $x^* \in \mathcal{X}$. For example, the following measure is defined by localization = ``point'', functional characteristic = ``function'', and loss = ``absolute'':

\begin{equation}\label{eq:measexample5}
   |\hat f(x^*) - f(x^*)| 
\end{equation}

This measure compares the values of $\hat f$ and $f$ at the point $x^*$.

When using point-specific measures, it typically makes sense to consider multiple points $x^* \in \mathcal{X}$. These points could correspond, e.g., to different quantiles of the distribution $F_X$. Returning to the NHANES example from Section \ref{sec:intro}, it could be of interest to compare the estimated to the true BMI values at different quantiles of age. 

As mentioned above, when comparing the estimated curve to the ground truth, the primary focus is often on the region where the variable $X$ typically takes its values. In such cases, regional measures (e.g., with scope of aggregation = $[F_X^{-1}(0.05),F_X^{-1}(0.95)]$) can be used to aggregate the deviation of the estimate from the truth across that region. However, in other cases (such as when extrapolation is a key goal), the behavior of the curves in the boundaries of the range of $X$ becomes important. For these situations, point-specific measures can be particularly useful, with the points chosen from the tails of the distribution $F_X$.

\subsection{Examples from the literature} Our categorization of measures includes several measures that were already proposed in the literature. For example, \textcite{buchholz2014measure} (see also \cite{govindarajulu2007comparing}) discussed the following measure: 

\begin{equation}\label{eq:precis}
    \int_{F_X^{-1}(0.01)}^{F_X^{-1}(0.99)} | \hat f(x) - f(x) | \hat p(x) \D x
\end{equation}

In the terminology of our categorization, this is the measure with localization = ``range'', axis of aggregation =  ``Y'', type of aggregation = ``expectation over the precision of the estimator'', scope of aggregation = $[F_X^{-1}(0.01), F_X^{-1}(0.99)]$, functional characteristic = ``function'', and loss = ``absolute''. Here, $\hat p(x)$ denotes the estimate for the precision of the estimator $\hat f$ at point $x$ (typically defined via the inverse of the variance of the estimator, \cite{buchholz2014measure}). Note that \textcite{buchholz2014measure} proposed this measure in a slightly different context, namely for modeling time-varying effects in survival analysis with non-linear functional forms, but the measure can also be used in our context, i.e., for modeling the effects of predictors. The idea behind taking the expectation over the precision of the estimator is to give more weight to the areas where the estimation is more precise (i.e., where the standard error of $\hat{f}(x)$ is small). In case of cubic regression splines with pre-specified knot positions, the precision of $\hat{f}(x)$ could, e.g., be obtained by evaluating the inverse of the expression $B(x) \hat{\Sigma} B(x)^\top$, where $B(x)$ is the row vector of basis functions (evaluated at $x$) and $\hat{\Sigma}$ is the estimated covariance matrix of the respective spline coefficient estimates (obtained from standard maximum likelihood fitting of an additive model, see \cite{harrell_regression_2015}). In case of P-splines, an estimate of the precision of $\hat{f}(x)$ could be derived analogously, taking a Bayesian view of the smoothing process and using the covariance of the posterior distribution of the spline coefficients (see Section 6.10 of \cite{wood2017gam}). The precision of fractional polynomials could be evaluated using bootstrap methods (see, e.g., \cite{royston2009bootstrap}).

Measures such as (\ref{eq:measexample}), (\ref{eq:measexample2}), (\ref{eq:measexample5}) and (\ref{eq:precis}) are defined with functional characteristic = ``function''. In some cases, the values $\hat f(x)$ of the estimated curve  might be quite close to the values $f(x)$ of the true curve for most $x$ values, but the curves still have rather different shapes. This will not be detected by measures with functional characteristic = ``function'', but could be detected by measures which consider the first and/or second derivatives of the curves. \textcite{binder2011multivariable} therefore proposed the following measure:

\begin{equation}\label{eq:ped1}
    \int_{F_X^{-1}(0.05)}^{F_X^{-1}(0.95)} \left(\hat f'(x) - f'(x)\right)^2 \D F_X(x)
\end{equation}

This is the measure with localization = ``range'', axis of aggregation = ``Y'', type of aggregation = ``expectation over $\D F_X$'', scope of aggregation = $[F_X^{-1}(0.05), F_X^{-1}(0.95)]$, functional characteristic = ``first derivative'', and loss = ``squared''. The measure evaluates how similar the slopes of the two curves are, with larger differences being punished by the squared loss. An analogous measure could be defined for functional characteristic = ``second derivative'' to compare the amount of ``wiggliness'' of the two curves. 

\subsection{Formulas for all measures} The formulas for all performance measures considered in this paper are given in Table \ref{tab:perf_desc_globalY} (measures with localization = ``range'' and axis of aggregation = ``Y''), Table \ref{tab:perf_desc_globalX} (measures with localization = ``range'' and axis of aggregation = ``X'') and Table \ref{tab:perf_desc_local} (measures with localization = ``point''). Formulas are given only for functional characteristic = ``function''; the analogous formulas for the derivatives can be obtained by replacing $f, \hat f$ with the respective derivatives. For the tables, some further notation is required.
We use the symbol $\mathcal{S}$ for the scope of aggregation, i.e., $\mathcal{S} \subseteq \mathcal{X}$ where either i) $\mathcal{S} = \mathcal{X}$ (if the scope of aggregation is the whole range of $X$) or ii) $\mathcal{S} = [F_X^{-1}(l), F_X^{-1}(u)]$ (if the scope of aggregation is between the $l$'th und $u$'th quantile of $F_X$). The definition of $\mathcal{S}$ could also be generalized to denote an arbitrary subset of $\mathcal{X}$. 

\renewcommand{\arraystretch}{1.5}
\begin{table}[h!]
\begin{small}
\begin{tabular}{p{4.5cm}ll}

  \textbf{Type of aggregation} & \textbf{Loss} & \textbf{Formula} \\\hline
   Integration over $\D x$ & Difference & $\int_{\mathcal{S}} (\hat f(x) - f(x)) \D x$ \\
    & Absolute & $\int_{\mathcal{S}} |\hat f(x) - f(x)| \D x$ \\
      & Squared & $\int_{\mathcal{S}} (\hat f(x) - f(x))^2 \D x$ \\
     & $\epsilon$-level accuracy &  $\int_{\mathcal{S}} \boldone_{\{|\hat f(x) - f(x)| \leq \epsilon\} } \D x $\\

   Expectation over $\D F_X(x)$ & Difference & $\int_{\mathcal{S}} (\hat f(x) - f(x)) \D F_X(x)$ \\
      & Absolute & $\int_{\mathcal{S}} |\hat f(x) - f(x)| \D F_X(x)$\\
      & Squared & $\int_{\mathcal{S}} (\hat f(x) - f(x))^2 \D F_X(x)$\\
      & $\epsilon$-level accuracy & $\int_{\mathcal{S}} \boldone_{\{|\hat f(x) - f(x)| \leq \epsilon\} } \D F_X(x) $\\

Quantile with respect to $F_X$ (e.g., median: $q = 0.5$) & Difference & $F^{-1}_{\hat f(X_{\mathcal{S}}) - f(X_{\mathcal{S}})} (q)$\\
      & Absolute & $F^{-1}_{|\hat f(X_{\mathcal{S}}) - f(X_{\mathcal{S}})|} (q)$\\
      & Squared & $F^{-1}_{(\hat f(X_{\mathcal{S}}) - f(X_{\mathcal{S}}))^2} (q)$ \\
     & $\epsilon$-level accuracy & $F^{-1}_{\boldone_{\{|\hat f(X_{\mathcal{S}}) - f(X_{\mathcal{S}})| \leq \epsilon\} }} (q)$ \\

   Expectation over precision of the estimator & Difference & $\int_{\mathcal{S}} (\hat f(x) - f(x)) \hat p(x) \D x$ \\
     & Absolute & $\int_{\mathcal{S}} |\hat f(x) - f(x)| \hat p(x) \D x$ \\
      & Squared & $\int_{\mathcal{S}} (\hat f(x) - f(x))^2 \hat p(x) \D x$ \\
      & $\epsilon$-level accuracy &  $\int_{\mathcal{S}} \boldone_{\{|\hat f(x) - f(x)| \leq \epsilon\} } \hat p(x) \D x $\\

   Maximum & Difference & $\max_{x \in \mathcal{S}} (\hat f(x) - f(x))$\\
      & Absolute & $\max_{x \in \mathcal{S}} |\hat f(x) - f(x)|$\\
      & Squared & $\max_{x \in \mathcal{S}} (\hat f(x) - f(x))^2$\\
      & $\epsilon$-level accuracy & $\max_{x \in \mathcal{S}} \boldone_{\{|\hat f(x) - f(x)| \leq \epsilon\} }$\\

   Minimum & Difference & $\min_{x \in \mathcal{S}} (\hat f(x) - f(x))$\\
      & Absolute & $\min_{x \in \mathcal{S}} |\hat f(x) - f(x)|$\\
      & Squared & $\min_{x \in \mathcal{S}} (\hat f(x) - f(x))^2$\\
      & $\epsilon$-level accuracy & $\min_{x \in \mathcal{S}} \boldone_{\{|\hat f(x) - f(x)| \leq \epsilon\} }$\\

\end{tabular}
 \caption{Performance measures for localization = ``range'' and axis of aggregation = ``Y''. The symbol $\mathcal{S}$ is a placeholder either for $\mathcal{S} = \mathcal{X}$ or $\mathcal{S} = [F_X^{-1}(l), F_X^{-1}(u)]$. Formulas are given for functional characteristic = ``function'', analogous formulas for $f'$, $f''$ can be obtained by replacing $f, \hat f$ with the respective derivatives.}
        \label{tab:perf_desc_globalY}
\end{small}
        \end{table}
\renewcommand{\arraystretch}{1}

\renewcommand{\arraystretch}{1.5}
\begin{table}[h!]
\begin{small}
\begin{tabular}{lll}

  \textbf{Type of aggregation} & \textbf{Loss} & \textbf{Formula} \\\hline
 Number of roots & Difference & $\mathrm{\# roots}^{\mathcal{S}}(\hat f) - \mathrm{\# roots}^{\mathcal{S}}(f)$ \\
      & Absolute & $|\mathrm{\# roots}^{\mathcal{S}}(\hat f) - \mathrm{\# roots}^{\mathcal{S}}(f)|$\\
      & Squared & $(\mathrm{\# roots}^{\mathcal{S}}(\hat f) - \mathrm{\# roots}^{\mathcal{S}}(f))^2$\\
      & $\epsilon$-level accuracy & $\boldone_{\{|\mathrm{\# roots}^{\mathcal{S}}(\hat f) - \mathrm{\# roots}^{\mathcal{S}}(f)| \leq \epsilon\}}$\\

   Location of maximum & Difference & $x^{\hat f, \mathrm{max}_\mathcal{S}} - x^{f, \mathrm{max}_\mathcal{S}}$\\
      & Absolute & $|x^{\hat f, \mathrm{max}_\mathcal{S}} - x^{f, \mathrm{max}_\mathcal{S}}|$ \\
      & Squared & $(x^{\hat f, \mathrm{max}_\mathcal{S}} - x^{f, \mathrm{max}_\mathcal{S}})^2$\\
     & $\epsilon$-level accuracy & $\boldone_{\{|x^{\hat f, \mathrm{max}_\mathcal{S}} - x^{f, \mathrm{max}_\mathcal{S}}| \leq \epsilon\}}$\\

   Location of minimum & Difference & $x^{\hat f, \mathrm{min}_\mathcal{S}} - x^{f, \mathrm{min}_\mathcal{S}}$\\
      & Absolute & $|x^{\hat f, \mathrm{min}_\mathcal{S}} - x^{f, \mathrm{min}_\mathcal{S}}|$ \\
      & Squared & $(x^{\hat f, \mathrm{min}_\mathcal{S}} - x^{f, \mathrm{min}_\mathcal{S}})^2$\\
     & $\epsilon$-level accuracy & $\boldone_{\{|x^{\hat f, \mathrm{min}_\mathcal{S}} - x^{f, \mathrm{min}_\mathcal{S}}| \leq \epsilon\}}$\\

\end{tabular}
 \caption{Performance measures for localization = ``range'' and axis of aggregation = ``X''. The symbol $\mathcal{S}$ is a placeholder either for $\mathcal{S} = \mathcal{X}$ or $\mathcal{S} = [F_X^{-1}(l), F_X^{-1}(u)]$. Formulas are given for functional characteristic = ``function'', analogous formulas for $f'$, $f''$ can be obtained by replacing $f, \hat f$ with the respective derivatives.}
        \label{tab:perf_desc_globalX}
\end{small}
        \end{table}
\renewcommand{\arraystretch}{1}

\renewcommand{\arraystretch}{1.5}
\begin{table}[h!]
\begin{small}
\begin{tabular}{ll}
   & \textbf{Loss} \\\hline
    Difference & $\hat f(x^*) - f(x^*)$\\
    Absolute & $|\hat f(x^*) - f(x^*)|$ \\
      Squared & $(\hat f(x^*) - f(x^*))^2$ \\
     $\epsilon$-level accuracy & $\boldone_{\{|\hat f(x^*) - f(x^*)| \leq \epsilon\} }$ \\

\end{tabular}
 \caption{Performance measures for localization = ``point'' (evaluation at a specific $x^* \in \mathcal{X}$). Formulas are given for functional characteristic = ``function'', analogous formulas for $f'$, $f''$ can be obtained by replacing $f, \hat f$ with the respective derivatives.}
        \label{tab:perf_desc_local}
\end{small}
        \end{table}
\renewcommand{\arraystretch}{1}

For measures with type of aggregation = ``quantile with respect to $F_X$'', let $F_{g(X_{\mathcal{S}})}$ denote the distribution of a random variable $g(X_{\mathcal{S}})$ that arises by sampling $X$ values from $P_X$ and discarding values outside of $\mathcal{S}$ (i.e., sampling from the distribution truncated to $\mathcal{S}$), and then applying some transformation $g$ to the remaining values (e.g., $g(x) = |\hat f(x) - f(x)|$). The corresponding quantile function is denoted as $F^{-1}_{g(X_{\mathcal{S}})}$. For example, this notation allows us to express the performance measure with localization = ``range'', axis of aggregation = ``Y'', type of aggregation = ``median with respect to $F_X$'', scope of aggregation = $\mathcal{S} = [F_X^{-1}(l), F_X^{-1}(u)]$, functional characteristic = ``function'', and loss = ``absolute'' with the following formula:
$$F^{-1}_{|\hat f(X_{\mathcal{S}}) - f(X_{\mathcal{S}})|}(0.5)$$ 
That is, values of $X$ are sampled between $F_X^{-1}(l)$ and $F_X^{-1}(u)$, the absolute differences $|\hat f(x) - f(x)|$ for all sampled values $x$ are calculated, and the median of these absolute differences is determined.

For measures with axis of aggregation = ``X'', we define the following quantities:

Location of maxima of $f, \hat f$ on $\mathcal{S}$:
$$x^{f, \mathrm{max}_{\mathcal{S}}} = \argmax_{x \in \mathcal{S}} f(x), ~x^{\hat f, \mathrm{max}_{\mathcal{S}}} = \argmax_{x \in \mathcal{S}} \hat f(x)$$
Location of minima of $f, \hat f$ on $\mathcal{S}$:
$$x^{f, \mathrm{min}_{\mathcal{S}}} = \argmin_{x \in \mathcal{S}} f(x), ~x^{\hat f, \mathrm{min}_{\mathcal{S}}} = \argmin_{x \in \mathcal{S}} \hat f(x)$$
Number of roots of $f, \hat f$ on $\mathcal{S}$:
$$\mathrm{\# roots}^{\mathcal{S}}(f), \mathrm{\# roots}^{\mathcal{S}}(\hat f)$$

For Tables \ref{tab:perf_desc_globalY} and \ref{tab:perf_desc_globalX}, we generally assume that the respective values (integrals, optima, etc.) exist and are well-defined and finite. In practice, this might not always be the case. For example, global measures might be ill-defined if the curves or their derivatives tend towards infinity at the boundaries of the full range. Moreover, some estimators tend to be very unstable in the tails of the distribution $F_X$, inducing a risk that global performance measures will be excessively impacted by the values estimated in regions where there is little data. In these cases, it can help to switch from global to regional measures by restricting the scope of aggregation from the whole range $\mathcal{X}$ to the interval $[F_X^{-1}(l), F_X^{-1}(u)]$ with suitable quantiles $l$, $u$.

%The measures in Table \ref{tab:perf_desc_globalY} also include monotonicity. For example, suppose we know that $\delta_1 := \min_{x \in \mathcal{X}} f'(x) > 0$, and let $\delta_2 := \min_{x \in \mathcal{X}} (\hat f'(x) - f'(x))$.
%If this minimum is less than zero, there is the possibility that $\min_{x \in \mathcal{X}} \hat f'(x)$ might not be $> 0$. 
%Then for all $x \in \mathcal{X}$, $\hat f'(x) \geq \delta_2 + f'(x) \geq \delta_2 + \delta_1$. This term is $> 0$ if $\delta_2 > - \delta_1$. However, this condition is not necessary. 

\subsection{Using performance measures in applied studies} Performance measures for estimated non-linear associations are not only relevant in simulation studies with a known ground truth, but could also be used as similarity measures in application studies to compare two curves estimated by two different methods. In this case, one can replace $f, \hat f$ in the formulas in Tables \ref{tab:perf_desc_globalY}-\ref{tab:perf_desc_local} with $\hat f^{(1)}, \hat f^{(2)}$, where $\hat f^{(1)}$ denotes the curve estimated by the first method and $\hat f^{(2)}$ the curve estimated by the second method. The precision $\hat p(x)$ then refers to the precision of the estimator of the difference $\hat f^{(1)}(x) - \hat f^{(2)}(x)$ (\cite{buchholz2014measure}). Note, however, that measures with difference loss are not symmetric with respect to the two estimated curves. %For example, the integrals $\int_{\mathcal{X}} (\hat f_1(x) - \hat f_2(x)) \D x$ and $\int_{\mathcal{X}} (\hat f_2(x) - \hat f_1(x)) \D x$ will generally not yield the same value. 
Mostly, these measures can be symmetrized by taking the absolute value of the measure, with the exception of measures with localization = ``range'' and axis of aggregation = ``Y'' where the type of aggregation is either the maximum or minimum. In this case, both the maximum and minimum difference loss should be considered.

\subsection{Calculating the measures} When calculating the measures with statistical software, some computational considerations have to be taken into account. Derivatives can sometimes be determined analytically, e.g., for (fractional) polynomials. For splines, the choice between calculating derivatives analytically or numerically depends on the specific type of spline and the software implementation. Point-specific measures are generally easier to compute than measures with localization = ``range''. For measures that involve integrals, the integrals can be approximated with Riemann sums. Note that some authors (\cite{strasak2011comparing,govindarajulu2007comparing}) present formulas for this type of measure directly as Riemann sums, not as integrals. Measures with the type of aggregation = ``expectation over $\D F_X$'' can be approximated either by using the density $\frac{d}{dx} F_X(x)$ of $X$ in combination with an approximation by Riemann sums, or by sampling i.i.d.\ observations $x$ from $P_X$ and taking the mean over the loss values, e.g., the mean over the differences $|\hat f(x) - f(x)|$ in measure (\ref{eq:measexample2}) (\cite{binder2011multivariable}).

For measures with localization = ``range'', axis of aggregation = ``Y'' and type of aggregation = ``maximum/minimum'', the maximum/minimum can be determined with a grid search.
For measures with localization = ``range'', axis of aggregation = ``X'' and type of aggregation = ``location of maximum/minimum'', the locations of the maximum/minimum can also be determined with a grid search. 

\section{Illustration of the behavior of different performance measures}
\label{sec:shinyapp}

In this section, we illustrate how different performance measures capture different properties of estimated curves. For illustration, we use several examples of ground truths and estimates that are available in a Shiny app at \url{https://thullmann.shinyapps.io/presplinesshinyapp/}. The selection of ground truths was inspired by applications in medical research, considering as wide a range of archetypical forms as possible. In the app, users can select a performance measure by choosing a specific option for each aspect of the categorization. The values of the chosen performance measure are then shown for four different examples. Each example consists of a ground truth curve $f$ and five different hypothetical estimates $\hat f$. The first estimate is always a linear fit. The other, non-linear estimates were derived from hand-drawn curves, representing various shapes where determining the optimal curve is not immediately obvious.

All curves and estimates are defined on the interval $\mathcal{X} = [0,1]$. Three different options for the distribution $F_X$ are available: $\textrm{Beta}(2,2)$, $\textrm{Beta}(2,5)$, and $\textrm{Beta}(5,2)$. For measures with localization = ``range'', there are two options for the scope of aggregation: the full range $\mathcal{X} = [0,1]$ or the subrange $[F_X^{-1}(0.05), F_X^{-1}(0.95)]$. For measures with loss = ``$\epsilon$-level accuracy'', different values for $\epsilon$ can be chosen. For measures with type of aggregration = ``quantile with respect to $F_X$'', different quantiles can be chosen as well. Measures with type of aggregation = ``expectation over precision of the estimator'' are not available in the app due to the manual definition of the hypothetical estimates. 

The code for the Shiny app is openly available on GitHub at \url{https://github.com/thullmann/performance-meas}. It includes an implementation of the performance measures from scratch in the statistical software R. 

\begin{figure}
    \centering    \includegraphics[width=0.7\textwidth]{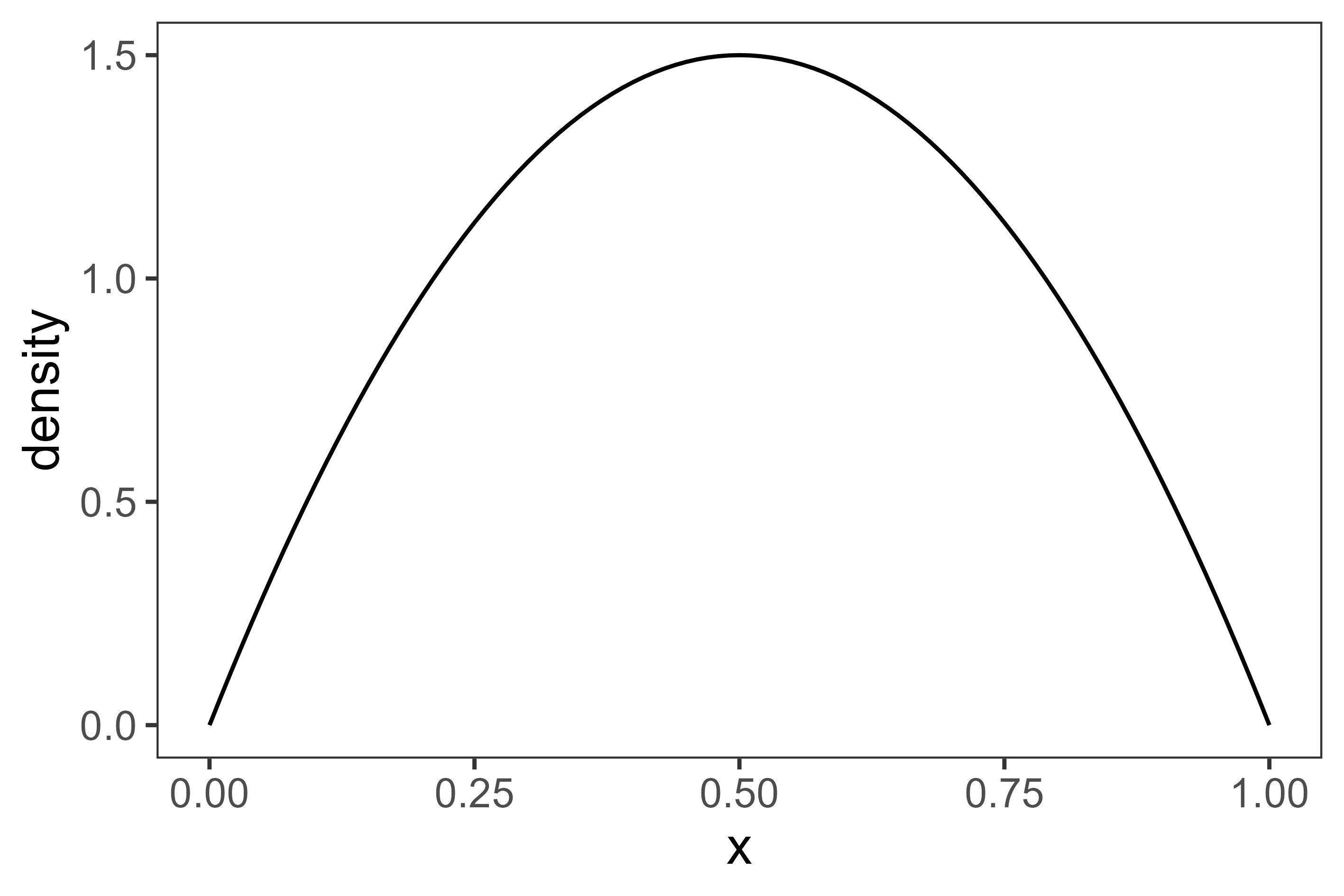}
    \caption{Density of the $\mathrm{Beta}(2,2)$ distribution.}
    \label{fig:fig4}
\end{figure}

In the following, we use three examples from the app and a subset of eleven measures from the categorization in Section \ref{sec:categ}. $F_X$ is set to $\textrm{Beta}(2,2)$. The density is shown in Fig~\ref{fig:fig4}. 

\subsection{Measures evaluating the function, the first derivative, or the second derivative}

\begin{figure}[h!]
    \centering    
    \includegraphics{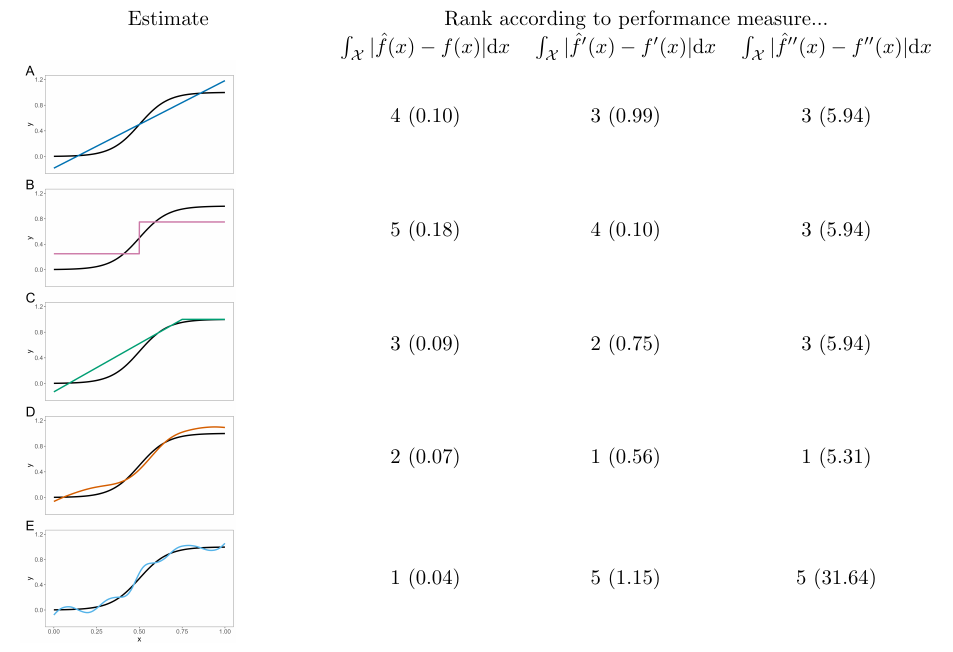}
\caption{Example: comparing measures that differ with respect to the functional characteristic (function, first derivative, or second derivative). The black curve is the true curve, the colored lines are five different ``estimates''. Next to each estimate, its rank among all five estimates according to the respective performance measure is shown. The values of the performance measures are shown in brackets, rounded to two decimal places.}
\label{fig:fig5}
\end{figure}

Fig~\ref{fig:fig5} shows an example from the Shiny app with three measures, all with localization = ``range'', axis of aggregation = ``Y'', type of aggregation = ``integral over $\D x$'', scope of aggregation = $\mathcal{X}$, and loss = ``absolute''. The measures differ with respect to the functional characteristic (function, first derivative, and second derivative). Next to each hypothetical estimate, its rank among the five estimates according to the respective performance measure is given. For the measure with functional characteristic = ``function'' (leftmost column in Fig~\ref{fig:fig5}), the most wiggly estimate E is ranked as the best one, because it is rather close to the true curve. However, when switching to the first or second derivative, estimate E is ranked as the worst estimate, because the slope and the wiggliness of the estimated curve are very different from the true curve, which is penalized by the respective measures.
In Appendix \ref{sec:screenshots}, screenshots of the Shiny app demonstrate how the first two measures can be selected in the app, yielding the ranks given in Fig~\ref{fig:fig5}. 

\subsection{Global versus regional measures}

\begin{figure}[h!]
       \centering    
       \includegraphics{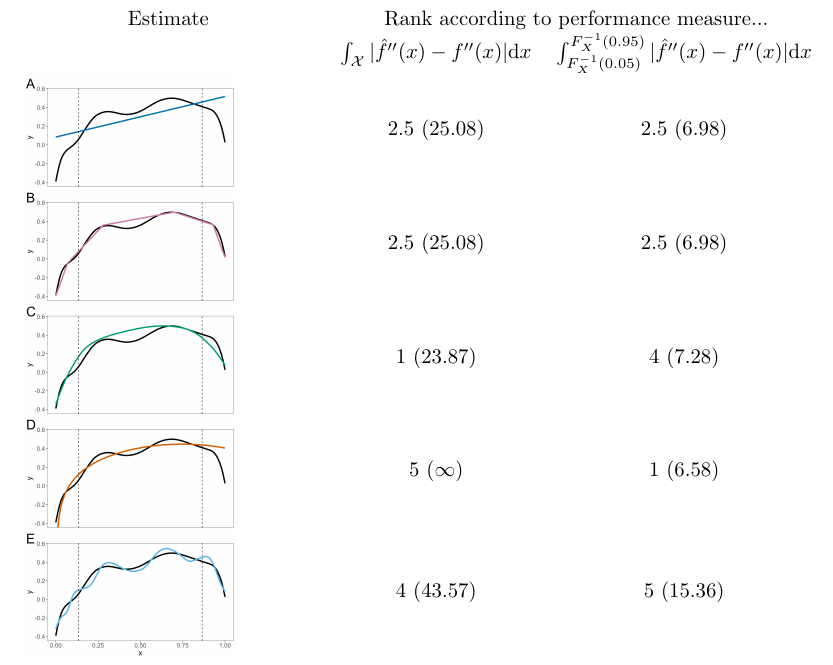}
    \caption{Example: comparing measures where the scope of aggregation is either the whole range $[0,1]$ or the subrange $[F_X^{-1}(0.05), F_X^{-1}(0.95)]$. The quantiles $F_X^{-1}(0.05)$ and $F_X^{-1}(0.95)$ are indicated by dashed vertical lines. The black curve is the true curve, the colored lines are five different ``estimates''. Next to each estimate, its rank among all five estimates according to the respective performance measure is shown. The values of the performance measures are shown in brackets, rounded to two decimal places.}
\label{fig:fig6}
\end{figure}

Fig~\ref{fig:fig6} shows another example from the app. Two different measures are compared, both with localization = ``range'', axis of aggregation = ``Y'', type of aggregation = ``integral over $\D x$'', functional characteristic = ``second derivative'' and loss = ``absolute''. The first measure uses the full range $\mathcal{X} = [0,1]$ as the scope of aggregation, the second measure the subrange $[F_X^{-1}(0.05), F_X^{-1}(0.95)]$. According to the first measure, estimate C is the best one. Estimate D is ranked last because the integral diverges, i.e., the performance value takes the value $\infty$ (due to the second derivative of estimate D tending to $\infty$ at a fast rate as $x \rightarrow 0$). However, according to the second measure, estimate C has only rank 4 and estimate D is the best one (because now the area around zero where the second derivative tends to $-\infty$ is excluded).    

\begin{figure}[h!]
        \centering   
        \includegraphics{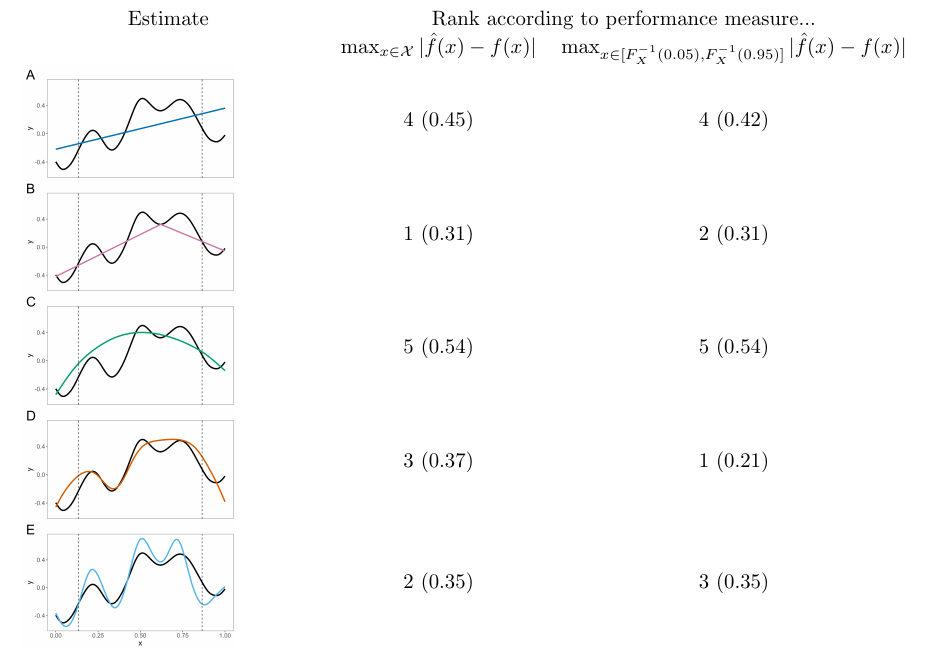}

    \caption{Example: comparing measures where the scope of aggregation is either the whole range $[0,1]$ or the subrange $[F_X^{-1}(0.05), F_X^{-1}(0.95)]$. The quantiles $F_X^{-1}(0.05)$ and $F_X^{-1}(0.95)$ are indicated by dashed vertical lines. The black curve is the true curve, the colored lines are five different ``estimates''. Next to each estimate, its rank among all five estimates according to the respective performance measure is shown. The values of the performance measures are shown in brackets, rounded to two decimal places.}
\label{fig:fig7}
\end{figure}

Fig~\ref{fig:fig7} shows a further example where different scopes of aggregation make a difference. Two different measures are compared, both with localization = ``range'', axis of aggregation = ``Y'', type of aggregation = ``maximum'', functional characteristic = ``function'' and loss = ``absolute''. As before, the first measure uses the full range $\mathcal{X} = [0,1]$ as the scope of aggregation, the second measure the subrange $[F_X^{-1}(0.05), F_X^{-1}(0.95)]$, which excludes both tails of the distribution. According to the first measure, estimate B (the piecewise linear fit) is the best, while estimate D has rank 3.  When restricting the scope of aggregation to the subrange $[F_X^{-1}(0.05), F_X^{-1}(0.95)]$, estimate B is ranked as second best and estimate D becomes the best one, because the areas towards the end points of the interval where estimate D has larger distances to the ground truth are excluded. 

\subsection{Measures with different types of aggregation: integral over $\D x$ versus the expectation over $\D F_X$}
\begin{figure}[h!]
    \centering    
    \includegraphics{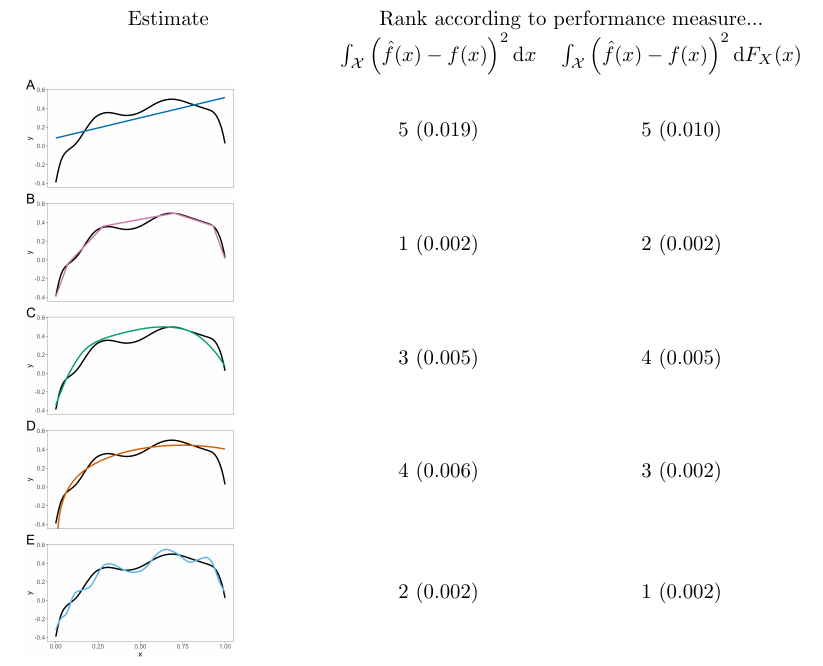}

    \caption{Example: comparing measures that use the integral over $\D x$ versus\ the expectation over $\D F_X$ for aggregating the loss. The black curve is the true curve, the colored lines are five different ``estimates''. Next to each estimate, its rank among all five estimates according to the respective performance measure is shown. The values of the performance measures are shown in brackets, rounded to three decimal places.}
\label{fig:fig8}
\end{figure}

Fig~\ref{fig:fig8} compares two different measures, both with localization = ``range'', axis of aggregation = ``Y'', scope of aggregation = $\mathcal{X}$, functional characteristic = ``function'' and loss = ``squared''. The type of aggregation is either the standard integral or the expectation over $\D F_X$. For the standard integral, estimate B (the piecewise linear fit) is ranked better than estimate E. For the expectation over $\D F_X$, it is the other way around.  Estimates C and D also switch ranks. This is because the expectation over $\D F_X$ puts more weight on the area at the center of the interval (where the density of $F_X$ peaks, see Fig~\ref{fig:fig4}). At the center, estimates B and C have more distance to the ground truth compared to estimates E and D, respectively.

\subsection{Measures with different types of aggregation: integral over $\D x$ versus maximum}
\begin{figure}[h!]
    \centering    
    \includegraphics{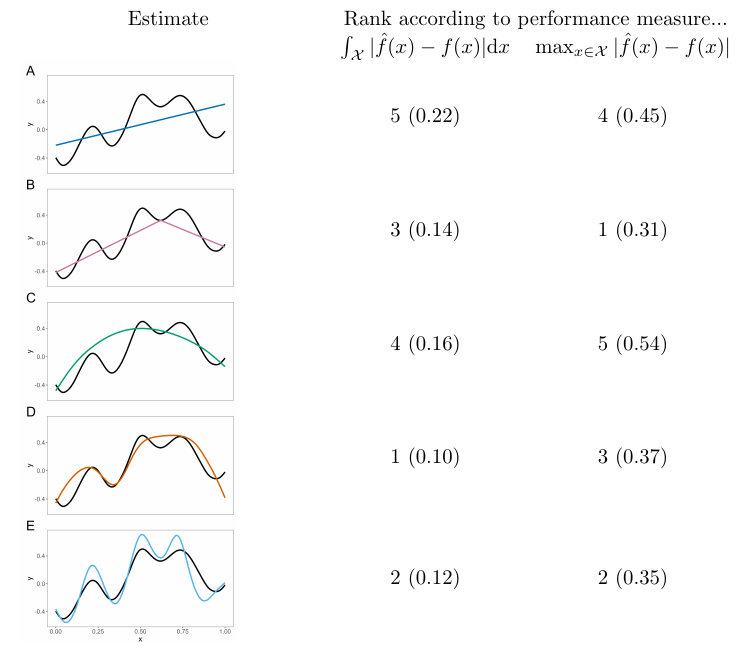}
    \caption{Example: comparing measures with the integrated loss versus the maximum loss as the type of aggregation. The black curve is the true curve, the colored lines are five different ``estimates''. Next to each estimate, its rank among all five estimates according to the respective performance measure is shown. The values of the performance measures are shown in brackets, rounded to two decimal places.}
\label{fig:fig9}
\end{figure}

Fig~\ref{fig:fig9} again compares different types of loss aggregation, now the integral over $\D x$ versus the maximum. The measures have localization = ``range'', axis of aggregation = ``Y'', scope of aggregation = $\mathcal{X}$, functional characteristic = ``function'', and loss = ``absolute''. The first measure integrates the loss over $\mathcal{X}$. According to this measure, estimate B (the piecewise linear fit) has rank 3 and estimate D is the best estimate. The second measure considers the maximum loss. Here, estimate B is the best estimate and estimate D now has rank 3. This can be explained as follows: on average, estimate D is closer to the ground truth compared to estimate B, but estimate D has a relatively large difference from the ground truth at the point $x = 1$.  

\subsection{Measures with different types of losses}
\begin{figure}[h!]
        \centering    
    \includegraphics{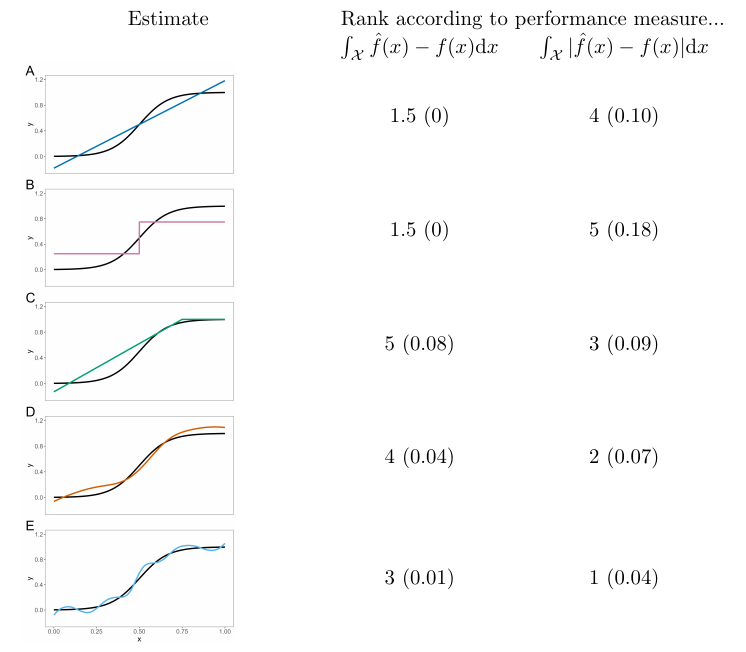}
 
    \caption{Example: comparing measures with difference loss versus absolute loss. The black curve is the true curve, the colored lines are five different ``estimates''. Next to each estimate, its rank among all five estimates according to the respective performance measure is shown. The values of the performance measures are shown in brackets, rounded to two decimal places.}
\label{fig:fig10}
\end{figure}

Different types of losses can lead to different rankings of estimates. Fig~\ref{fig:fig10} compares the difference loss to the absolute loss.  Both measures have localization = ``range'', axis of aggregation = ``Y'', type of aggregation = ``integral over $\D x$'', scope of aggregation = $\mathcal{X}$, and functional characteristic = ``function''. The first measure uses the difference loss, the second measure the absolute loss. With the difference loss, estimates A and B are best with an integrated bias of zero, because the positive and negative signs of the differences from the ground truth cancel each other out. However, according to the measure with the absolute loss, these estimates are the worst ones, because they are rather different from the ground truth in absolute terms.

\begin{figure}[h!]
    \centering    
    \includegraphics{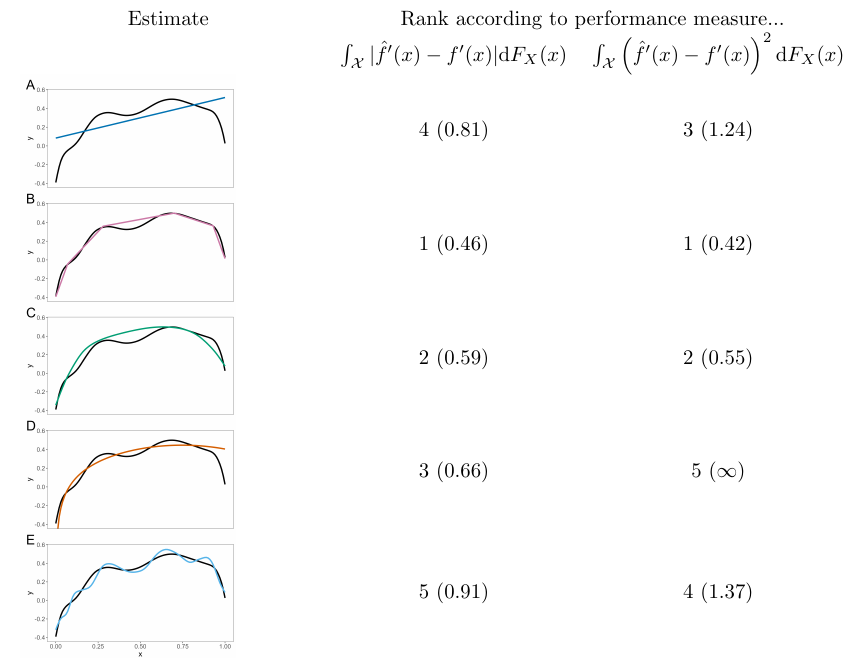}
 
    \caption{Example: comparing measures with absolute loss versus squared loss. The black curve is the true curve, the colored lines are five different ``estimates''. Next to each estimate, its rank among all five estimates according to the respective performance measure is shown. The values of the performance measures are shown in brackets, rounded to two decimal places.}
\label{fig:fig11}
\end{figure}

Fig~\ref{fig:fig11} shows an example for the absolute loss versus the squared loss. The two measures have localization = ``range'', axis of aggregation = ``Y'', type of aggregation = ``expectation over $\D F_X$'', scope of aggregation = $\mathcal{X}$, and functional characteristic = ``first derivative''. The first measure uses the absolute loss, the second measure the squared loss. With respect to the absolute loss, estimate D is ranked third, but only fifth with respect to the squared loss. The derivative of estimate D tends to $\infty$ for $x \rightarrow 0$, but at such a rate that the expectation over $\D F_X$ still converges when using the absolute loss. For the squared loss, the integral diverges to $\infty$, making estimate D the worst estimate.

In summary, the examples in Fig~\ref{fig:fig5}-Fig~\ref{fig:fig11} showed that different performance measures can rank estimates in different ways. Different measures emphasize different aspects of the estimated curves. However, this does not mean that different performance measures always result in entirely different rankings. As can be seen in the Shiny app, many measures are similar to each other with respect to the resulting rankings. We will return to this point in the Discussion.

\section{Discussion}
\label{sec:discuss}

In the literature, there is a lack of consensus on how to evaluate the accuracy of non-linear functional form estimates in additive regression. Here, we have suggested a categorization of performance
measures for estimated non-linear associations that could, e.g., be used to evaluate and compare fits obtained from different types of spline regression and/or fractional polynomial regression. 

\subsection{Performance measures for multivariable models}

We have considered only ``univariable'' performance measures, i.e., the estimated curve of each predictor of interest is evaluated individually, though the curve can be adjusted for the estimated effects of other predictors in a multivariable model. This approach is typical for descriptive modeling, where one is interested in correctly capturing the functional form of a predictor of interest. Although it could in principle be possible to aggregate individual performance values across predictors to obtain a single performance metric for the whole model, this only makes sense if the ranges of the performance measures are comparable across predictors with different distributions. Also, after aggregation, the quality of the individual curves cannot be assessed anymore. In contrast, calculating a single metric for the whole model is more relevant for predictive modeling where the prediction error is such a metric.  

The univariable performance measures we consider 
do not directly cover multivariable non-linear effects, e.g., when the effect of two variables $X_1, X_2$ is given in a non-linear form $f(X_1,X_2)$ that does not boil down to the transformation of a simple product variable $V$. However, our categorization of measures could be generalized to cover such cases. For example, the univariable measure
    $$\int_{\mathcal{X}} |\hat f(x) - f(x)| \D F_X(x)$$
can be generalized to the following bivariable measure for two variables $X_1, X_2$:
    $$\int_{\mathcal{X}_1 \times \mathcal{X}_2} |\hat f(x_1,x_2) - f(x_1,x_2)| \D F_{(X_1, X_2)}(x),$$

with $\mathcal{X}_1 \times \mathcal{X}_2 \subseteq \mathbb{R}^2$ denoting the space where $(X_1,X_2)$ take values and $F_{(X_1, X_2)}$ denoting the joint distribution of $(X_1, X_2)$. This integral can then be approximated with Riemann sums by partitioning $\mathcal{X}_1 \times \mathcal{X}_2$ into sub-rectangles.

\subsection{Performance measures for predictive and explanatory modeling}

As mentioned in the Introduction, we have considered performance measures in the context of descriptive modeling. Estimating non-linear functional forms can also be relevant for simulation studies on predictive or explanatory modeling, but in these contexts, the evaluation of the estimates might require somewhat different performance measures. In predictive modeling, one is typically less interested in what the functional form of a predictor looks like exactly, only in whether modeling the functional form in a suitable (potentially non-linear) way can improve prediction. The prediction error is then the most relevant performance measure. In explanatory modeling, where one aims to estimate the causal effect of an explanatory variable, we can distinguish between two cases: whether the explanatory variable is binary or continuous. If the explanatory variable is binary, non-linear functional forms for continuous confounders could be estimated in order to circumvent misspecification of the confounder-outcome relation and to avoid residual confounding (\cite{brenner1997controlling,benedetti2004using}). A too complex functional form may lead to increased variance of the intervention effect, while underfitting the confounder-outcome relation may lead to bias in the  effect of the explanatory variable of interest (but probably a lower variance). Hence, the accuracy of the estimated effect measured, e.g., with the root mean squared error of the respective regression coefficient, is a relevant performance measure. If the explanatory variable is continuous (e.g., when estimating a dose-response relationship between an exposure and an outcome), non-linear functional forms could again be estimated for continuous confounders, but in addition, the dose-response relation for the exposure could also be modeled non-linearly (\cite{desquilbet2010dose}). In that case, the performance measures in the present paper could be used to compare the estimated dose-response curve to the true curve. Exploring performance measures in the context of predictive and explanatory modeling in more detail goes beyond the scope of the present article but could be addressed in future research.

\subsection{Choosing a suitable subset of performance measures}

Our categorization of performance measures includes a large number of measures. If we consider one option for the type of aggregation = ``quantile with respect to $F_X$'' (e.g., the median) and two options for the scope of aggregation ($\mathcal{X}$ and $[F_X^{-1}(l), F_X^{-1}(u)]$ for some specific quantiles $l, u$, for example $l = 5\%$, $u = 95\%$), there are 216 measures with localization = ``range''. Additionally, for each point $x^* \in \mathcal{X}$, there are 12 measures with localization = ``point''. Obviously, not all of these performance measures should be included in a simulation study which compares different methods for estimating non-linear associations. At least, not all measures can be reported in the main manuscript. The question then is, how researchers can choose a smaller, sufficient set of performance measures for a simulation study. A pre-selection of measures might possibly be based on the simulation settings: the ``ground truth'' curves might have certain properties that make some measures more relevant than others. Yet, the set of possible measures might still be rather large after this pre-selection. To further reduce the set of measures, one could look for groups of measures that give similar results, i.e., that attribute similar relative performance to the methods. Such groups could be detected, e.g., with a clustering algorithm. Then it would suffice to report only one measure per group. We plan to explore this idea in future research.

\section*{Funding Information}

This work was supported through the Austrian Science Fund (FWF) project I-4739-B to DD and the German Research Foundation (DFG) grant SA580/10-3 to WS.

\section*{Data availability}

All data and code are openly available on GitHub at \url{https://github.com/thullmann/performance-meas}.

\printbibliography

\newpage
\appendix

\section{Shiny app screenshots}
\label{sec:screenshots}

    \begin{figure}[h]
    \centering
    \includegraphics[width=\textwidth]{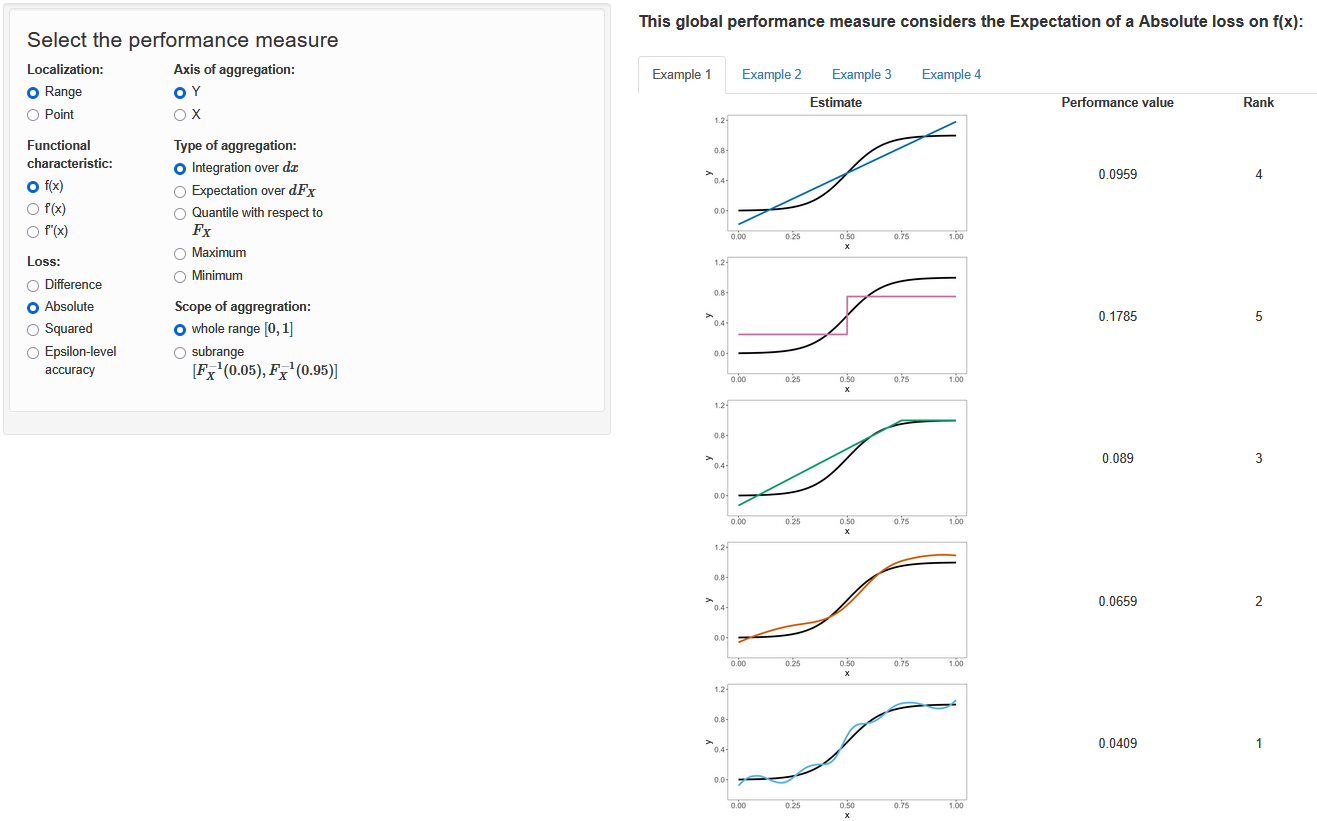}
    \caption{This screenshot of the Shiny app demonstrates how to reproduce the ranks for the performance measure $\int_{\mathcal{X}} |\hat f(x) - f(x)| \D x$ given in Fig~5 in the paper. The measure can be selected with the radio buttons on the left hand side by choosing localization = ``range'', functional characteristic = ``$f(x)$'', loss = ``absolute'', axis of aggregation = ``Y'', type of aggregation = ``integral over $\D x$'', and scope of aggregation = ``whole range $[0,1]$''.}
    \label{fig:screenshot1}
\end{figure}

    \begin{figure}[h]
    \centering
    \includegraphics[width=\textwidth]{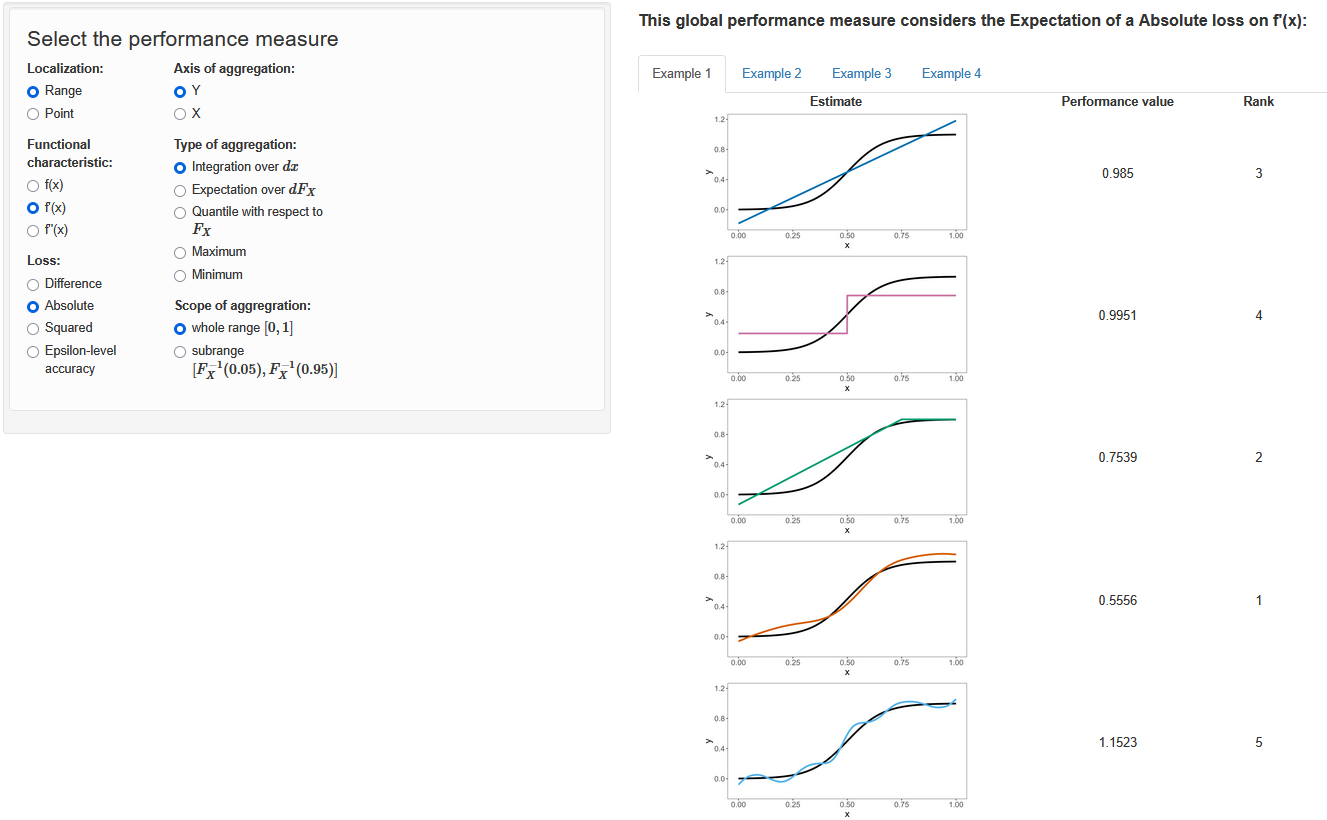}
    \caption{This screenshot of the Shiny app demonstrates how to reproduce the ranks for the performance measure $\int_{\mathcal{X}} |\hat f'(x) - f'(x)| \D x$ given in Fig~5 in the paper. The measure can be selected with the radio buttons on the left hand side by choosing localization = ``range'', functional characteristic = ``$f'(x)$'', loss = ``absolute'', axis of aggregation = ``Y'', type of aggregation = ``integral over $\D x$'', and scope of aggregation = ``whole range $[0,1]$''. That is, the only difference to the selection in Fig~\ref{fig:screenshot1} concerns the functional characteristic.}
    \label{fig:screenshot2}
\end{figure}

\end{document}